\documentclass[prd,preprint,superscriptaddress,showpacs,nofootinbib,%
tightenlines
]{revtex4}
\newcommand{\ben}{\begin{displaymath}}
\newcommand{\een}{\end{displaymath}}
\newcommand{\be}{\begin{equation}}
\newcommand{\ee}{\end{equation}}
\newcommand{\bea}{\begin{eqnarray}}
\newcommand{\eea}{\end{eqnarray}}

\begin{document}
\preprint{MKPH-T-03-13}
\title{Infrared regularization of baryon chiral perturbation theory
reformulated}
\author{Matthias R.~Schindler}
\affiliation{Institut f\"ur Kernphysik, Johannes
Gutenberg-Universit\"at, D-55099 Mainz, Germany}
\author{Jambul Gegelia}
\thanks{Alexander von Humboldt Research Fellow}
\affiliation{Institut f\"ur Kernphysik, Johannes
Gutenberg-Universit\"at, D-55099 Mainz, Germany}
\affiliation{High Energy Physics Institute,
Tbilisi State University,
University St.~9, 380086 Tbilisi, Georgia}
\author{Stefan Scherer}
\affiliation{Institut f\"ur Kernphysik, Johannes
Gutenberg-Universit\"at, D-55099 Mainz, Germany}
\begin{abstract}
   We formulate the infrared regularization of Becher and Leutwyler in a
form analogous to our recently proposed extended on-mass-shell
renormalization.
   In our formulation, IR regularization can be applied straightforwardly to
multi-loop diagrams with an arbitrary number of particles
with arbitrary masses.
\end{abstract}
\pacs{
11.10.Gh,
12.39.Fe.
}
\date{31 August, 2003}
\maketitle

\section{\label{introduction} Introduction}
   Starting with Weinberg's fundamental work on phenomenological Lagrangians
\cite{Weinberg:1979kz}, it became possible to systematically calculate
corrections to the soft-pion results obtained within the framework of
current algebra \cite{Adler:1968}.
   The corresponding effective field theory (EFT)---chiral perturbation
theory (ChPT)---has been very successful in describing the strong interactions
at low energies (for a recent review see, e.g., Ref.\ \cite{Scherer:2002tk}).
   In the mesonic sector, the combination of standard dimensional
regularization (DR) and the modified minimal subtraction scheme of ChPT
($\widetilde{\rm MS}$) led to a straightforward correspondence between the
loop expansion and the chiral expansion in terms of momenta and quark masses
at a fixed ratio \cite{Gasser:1984yg,Gasser:1984gg}.
   The one-baryon sector proved to be more complicated \cite{Gasser:1988rb}.
   In particular, using the same combination of DR and $\widetilde{\rm MS}$
as in mesonic ChPT, higher-order loops contribute in lower chiral orders
and therefore the correspondence between the loop expansion and the chiral
expansion seems to be lost (see Fig.\ 2 of Ref.\ \cite{Gasser:1988rb}).
    One solution to this power-counting problem was given in the framework of
heavy-baryon chiral perturbation theory (HBChPT) \cite{Jenkins:1991jv}, and
most of the recent calculations have been performed within this approach
\cite{Bernard:1992qa,Bernard:1995dp}.
   While successful in many cases, HBChPT destroys the ana\-lytic structure in
part of the low-energy region.
   Several methods have been suggested to reconcile power counting
with the constraints of analyticity in a manifestly relativistic approach
\cite{Ellis:1997kc,Becher:1999he,Gegelia:1999gf,Gegelia:1999qt,%
Lutz:1999yr,Fuchs:2003qc}.
   The one most widely used is the infrared (IR) regularization of
Ref.~\cite{Becher:1999he} by Becher and Leutwyler.
   A possible generalization to multi-loop diagrams has been suggested
in Ref.\ \cite{Lehmann:2001xm}.

    In the present paper we provide a formulation of the IR regularization
of Becher and Leutwyler in a form analogous to the extended on-mass-shell
(EOMS) renormalization of Ref.\ \cite{Fuchs:2003qc}.
   As a result of the reformulation, IR regularization is straightforwardly
applicable to multi-loop diagrams as well as to diagrams involving several
fermion lines and/or resonances.

\section{\label{comparison}Comparison of IR regularization and
EOMS renormalization}

   In order to reformulate the infrared regularization of Becher and
Leutwyler \cite{Becher:1999he} in a form analogous to the EOMS
renormalization of Ref.~\cite{Fuchs:2003qc}, we consider as an
example the characteristic, dimensionally regularized, one-loop
integral of the fermion self-energy,
\begin{equation}
\label{intex}
I_{N\pi}(-p,0)=i\int\frac{d^nk}{(2\pi)^n}\frac{1}{\left[(k-p)^2-m^2+i 0^+
\right]\left[ k^2-M^2+i 0^+ \right]},
\end{equation}
where $n$ denotes the number of space-time dimensions.
   The masses $m$ and $M$ refer to the nucleon mass in the chiral limit
and the lowest-order pion mass, respectively.
   Using the standard power counting of Refs.~\cite{Weinberg:um,Ecker:1994gg}
we assign the order $Q^{n-3}$ to the integral $I_{N\pi}$.
   Here, $Q$ collectively denotes small expansion parameters such as the pion
mass or small external momenta.
   (Note that $I_{N\pi}$ satisfies power counting only after
subtraction \cite{Becher:1999he,Fuchs:2003qc}).

   To implement the IR regularization and to compare with the EOMS
renormalization scheme we use the Feynman parametrization formula
\begin{equation}
\label{fpar}
\frac{1}{a b}=\int_0^1 \frac{d z}{\left[az+b(1-z)\right]^2},
\end{equation}
with $a=(k-p)^2-m^2+i 0^+$ and $b=k^2-M^2+i 0^+$, interchange the
order of integrations, perform the integration over loop momenta
$k$, and obtain
\begin{equation}
\label{intpar}
I_{N\pi}(-p,0)=-\frac{1}{(4\pi)^{n/2}} \ \Gamma(2-n/2) \int_0^1 dz
\ \left[ A(z)\right]^{(n/2)-2},
\end{equation}
where
$$A(z)=-p^2 (1-z) z+m^2 z+M^2 (1-z)-i 0^+.$$
   In the approach of Becher and Leutwyler, the integral $I_{N\pi}$ is
divided into the IR singular part $I$ and the remainder $R$,
$I_{N\pi}=I+R$, defined as
\begin{equation}
\label{I}
I=-\frac{1}{(4\pi)^{n/2}} \ \Gamma(2-n/2) \int_0^\infty dz \left[
A(z)\right]^{(n/2)-2},
\end{equation}
\begin{equation}
\label{R}
R=\frac{1}{(4\pi)^{n/2}} \ \Gamma(2-n/2) \int_1^\infty  dz \left[
A(z)\right]^{(n/2)-2}.
\end{equation}
  In this decomposition, for noninteger $n$ the integral $I$ is proportional
to a noninteger power of the pion mass ($\sim M^{n-3}$) and
thus satisfies the power counting.
   On the other hand, the remainder $R$ does not satisfy the power counting
but, for arbitrary $n$, contains non-negative powers of small parameters
and is thus absorbed into an infinite number of counterterms.
   Divergent parts of $I$ are also absorbed in an infinite number of
counter\-terms.

   In our EOMS renormalization scheme \cite{Fuchs:2003qc} we apply a
conventional renormalization prescription which allows us to
identify the terms which we subtract from a given integral
without calculating the integral beforehand.
   In essence we work with a modified integrand which is obtained from the
original integrand by subtracting a suitable number of counterterms.
   To find the subtraction terms we consider the series
\begin{eqnarray}
\lefteqn{\sum_{l,j=0}^\infty \frac{(p^2-m^2)^l (M^2)^j}{l!j!}}\nonumber\\
&&\times \left\{
\left(\frac{1}{2p^2}p_\mu\frac{\partial}{\partial p_\mu}\right)^l
\left(\frac{\partial}{\partial M^2}\right)^j
\frac{1}{[(k-p)^2-m^2+i0^+][k^2-M^2+i0^+]}\right\}_{p^2=m^2, M^2=0}
\nonumber\\
&=&\left.\frac{1}{\left( k^2-2 k\cdot p+i 0^+ \right)\left( k^2+i
0^+\right)}\right|_{p^2=m^2}
+M^2 \left. \frac{1}{\left( k^2-2 k\cdot p+i 0^+
\right)\left( k^2+i 0^+\right)^2 }\right|_{p^2=m^2}\nonumber\\
&& +(p^2-m^2)
\Biggl[ \frac{1}{2 m^2} \frac{1}{\left( k^2-2 k\cdot p+i 0^+
\right)^2}-\frac{1}{2 m^2} \frac{1}{\left( k^2-2 k\cdot p+i 0^+
\right)\left( k^2+i 0^+\right) } \nonumber\\
&&
\label{ss}
-\frac{1}{\left( k^2-2 k\cdot p+i 0^+ \right)^2\left(
k^2+i 0^+\right)}\Biggr]_{p^2=m^2}+\cdots,
\end{eqnarray}
    where $[\ldots]_{p^2=m^2}$ means that we consider the
coefficients of $(p^2-m^2)^l (M^2)^j$ only for four-momenta
$p^\mu$ which satisfy the on-mass-shell condition.
   Although the coefficients still depend on the direction of $p^\mu$, after
integration of this series with respect to the loop momenta $k$
and evaluation of the resulting coefficients for $p^2=m^2$, the
integrated series is a function of $p^2$ only.
   We subtract from Eq.~(\ref{intex}) those terms of the expansion
of Eq.\ (\ref{ss}) which violate the power counting.
   These terms are analytic in the small parameters and do not contain
infrared divergences.
   For the given example we only need to subtract the first term of the
expansion of Eq.~(\ref{ss}).

   We note that integrating Eq.~(\ref{ss}) term by term
reproduces the expansion of $R$ of Eq.\ (\ref{R}) in $M^2$ and $p^2-m^2$.
   This can be checked by explicitly integrating the first few coefficients of
the expansion of Eq.~(\ref{ss});
   we indeed see that they coincide with the
coefficients of the expansion of $R$ of Ref.~\cite{Becher:1999he}:
\begin{equation}
\label{Rrepr} R=-\frac{m^{n-4} \Gamma(2-n/2)}{(4\pi)^{n/2} (n-3)}
\left[1-\frac{p^2-m^2}{2 m^2}+\frac{(n-6)
\left(p^2-m^2\right)^2}{4 m^4 (n-5)}+\frac{(n-3) M^2 }{2 m^2 (n-5)
}+\cdots\right].
\end{equation}
   A more straightforward and transparent way of obtaining the IR regular
part $R$ is to rewrite $I_{N\pi}$ using the Feynman (or Schwinger)
parameterization, integrate over loop momenta, expand the resulting
integrand (of the integration over parameters) in a Taylor series
of Lorentz-invariant small expansion parameters (small masses and
Lorentz-invariant combinations of external momenta and large
masses), and, finally, interchange summation and integration:
$\int dx \sum\to \sum\int dx$.
   As is shown in the next section, the above observation is correct in
general, i.e., by expanding the integrand of any integral with an
arbitrary number of nucleon and pion denominators in small
parameters and interchanging summation and integration, one
reproduces the expansion of the IR regular part of the
integral.\footnote{Note the important difference with
Ref.~\cite{Ellis:1997kc}, where the expansion of the integrand
with subsequent interchange of integration and summation
reproduces the chiral expansion of the power-counting preserving
part.}

\section{General case}

   Let us consider the general one-loop scalar integral corresponding
to diagrams with one fermion line and an arbitrary number of pion
and fermion propagators:
\begin{equation}
\label{genint} I_{N\cdots\pi\cdots}(p_1,\ldots,q_1,\ldots)=i \int
\frac{d^nk}{(2\pi)^n} \ \frac{1}{b_1\cdots b_l \ a_1\cdots a_m},
\end{equation}
where $$b_j=(k+p_j)^2-m^2+i 0^+, $$ $$a_i=(k+q_i)^2-M^2+i 0^+.$$
   Tensor integrals are reduced to the scalar integrals of
Eq.~(\ref{genint}) in the standard fashion \cite{Passarino:1978jh}.

   Following Ref.~\cite{Becher:1999he} we apply the infrared
regularization to the integral of Eq.~(\ref{genint}).
   We start by combining all meson propagators using the formula
\begin{equation}
\label{fpargen} \frac{1}{a_1\cdots
a_m}=\left(\frac{\partial}{\partial M^2}\right)^{(m-1)} \int_0^1
dx_1\cdots \int_0^1 dx_{m-1}\frac{X}{A}.
\end{equation}
   The numerator $X$ is given by
$$
X=\left\{
\begin{array}{l}
1\,\,\mbox{for}\,\, m=2,\\
x_2 (x_3)^2\cdots (x_{m-1})^{m-2}\,\,\mbox{for}\,\,m> 2,
\end{array}
\right.
$$
and the denominator $A$ is
given by the recursive expression
\begin{eqnarray*}
A&=&A_m,\\
A_1&=&a_1,\\
A_{p+1}&=&x_p A_p+(1-x_p) a_{p+1} \ \ \ \ (p=1,\ldots ,m-1).\\
\end{eqnarray*}
   The result for $A$ is of the form
\begin{equation}
\label{adres} A=(k+\bar q)^2-\bar A+i 0^+,
\end{equation}
where the constant term $\bar A$ is of order $Q^2$, and $\bar q$ is a
linear combination of external momenta and is of order $Q^1$.

   Analogously we combine the nucleon propagators
\begin{equation}
\label{fpargenb} \frac{1}{b_1\cdots
b_l}=\left(\frac{\partial}{\partial m^2}\right)^{(l-1)} \int_0^1
dy_1\cdots \int_0^1 dy_{l-1}\frac{Y}{B}.
\end{equation}
   The numerator $Y$ is given by
$$
Y=\left\{
\begin{array}{l}
1\,\,\mbox{for}\,\,l=2,\\
y_2 (y_3)^2\cdots (y_{l-1})^{l-2}\,\,\mbox{for}\,\, l> 2,
\end{array}
\right.
$$
and the denominator $B$ is given by the recursive expression
\begin{eqnarray*}
B&=&B_l,\\
B_1&=&b_1,\\
B_{p+1}&=&y_p B_p+(1-y_p) b_{p+1} \ \ \ \ (p=1,\ldots ,l-1).\\
\end{eqnarray*}
The result for $B$ reads
\begin{equation}
\label{bdres} B=(k+\bar P)^2-\bar B+i 0^+,
\end{equation}
where $\bar P$ is a linear combination of external momenta, $\bar
P^2=m^2+{\cal O}(Q)$ and $\bar B=m^2+{\cal O}(Q)$.

   Next we combine the denominators $A$ and $B$ using
$$\frac{1}{A B}=\int_0^1 \frac{d z}{\left[ (1-z) A+z
B\right]^2}$$
and obtain for the integral of Eq.~(\ref{genint})
\begin{eqnarray}
\lefteqn{i \left(\frac{\partial}{\partial
M^2}\right)^{(m-1)}\left(\frac{\partial}{\partial
m^2}\right)^{(l-1)} \int_0^1 dz \int_0^1 dy_1\cdots \int_0^1
dy_{l-1}}\nonumber\\
&&\times
\label{intrepr}
\int_0^1 dx_1\cdots\int_0^1 dx_{m-1}\, Y X  \int
\frac{d^n k}{(2 \pi)^n}\frac{1}{\left[ (1-z) A+z B\right]^2}.
\end{eqnarray}
   Substituting $A$ and $B$ from Eqs.~(\ref{adres}) and (\ref{bdres})
in Eq.~(\ref{intrepr}), evaluating the derivatives, and shifting
$k\rightarrow k-\bar P z-\bar q (1-z)$, we obtain
\begin{equation}
\label{intrepr2}
i (l+m-1)! \int_0^1 dz\, z^{l-1} (1-z)^{m-1} \int_0^1
dy_1\cdots
\int_0^1 dx_{m-1}\, YX
\int \frac{d^n k}{(2 \pi)^n}\frac{1}{[k^2-f(z)]^{l+m}},
\end{equation}
where
$$
f(z)=\bar P^2 z^2-\left(\bar P^2-\bar B\right) z+ \bar A (1-z)
-\left( \bar q^2-2 \bar
P\cdot\bar q \right) z (1-z)-i 0^+.
$$
   Finally, the integration of Eq.~(\ref{intrepr2}) over $k$ yields
\begin{equation}
\label{intrepr3}
\frac{(-1)^{1-l-m}}{(4 \pi)^{n/2}} \ \Gamma(l+m-n/2)
\int_0^1 dz\,
z^{l-1} (1-z)^{m-1} \int_0^1 dy_1\cdots
\int_0^1 dx_{m-1}\, Y X [f(z)]^{(n/2)-l-m}.
\end{equation}
   To apply the IR regularization we rewrite the $z$ integration
as
$$
\int_0^1dz \cdots =\int_0^\infty dz \cdots-\int_1^\infty dz \cdots.
$$
    The result of the first integration is identified as the
IR singular part and of the second as the IR regular part. In
the IR regular part one can expand the integrand in small momenta
and masses and interchange summation and integration
\cite{Becher:1999he}.
   This leads to integrals over $z$ of the type
\begin{equation}
\label{Iiint} I_i=\int_1^\infty dz z^{n+i},
\end{equation}
where $i$ is an integer number.
   These $I_i$ are multiplied by (further) integrals over
$x_j$ and $y_k$ which do not depend on $n$.
   The integrals of Eq.~(\ref{Iiint}) are calculated by analytic
continuation from the domain of $n$ in which they converge, i.e.
\begin{equation}
\label{Iiintcalc} I_i=\left.\frac{z^{n+i+1}}{n+i+1}\right|_1^\infty
=-\frac{1}{n+i+1}.
\end{equation}

   On the other hand, if we expand the integrand in
Eq.~(\ref{intrepr3}) in small momenta and masses and interchange
summation and integration, we obtain exactly the same expansion as for the IR
regular part of the IR regularization with the only difference
that instead of the integrals $I_i$ of Eq.~(\ref{Iiint}) we now
have\footnote{The minus sign relative to Eq.\ (\ref{Iiint}) stems from
the definition of $R$ as $-\int_1^\infty dz\cdots$.}
\begin{equation}
\label{Jiint} J_i=-\int_0^1 dz z^{n+i}.
\end{equation}
   Calculating these integrals by analytical continuation from the
domain of $n$ in which they converge, we obtain:
\begin{equation}
\label{Jiintcalc}
J_i=-\left.\frac{z^{n+i+1}}{n+i+1}\right|_0^1=-\frac{1}{n+i+1}.
\end{equation}
   Comparing Eqs.~(\ref{Iiintcalc}) and (\ref{Jiintcalc}) we see that
the expansion of the integrand in Eq.~(\ref{intrepr3}) with
subsequent interchange of summation and integration exactly
reproduces the result of the IR regular part of the loop integral.
   Next we observe that, if we expand the integrand of
Eq.~(\ref{intrepr2}) in small parameters and interchange
summation and integration over $k$, we obtain exactly the same
result as by expanding the integrand in Eq.~(\ref{intrepr3}) in
small parameters with subsequent interchange of summation and
integration over Feynman parameters.
   We further note that the result of the expansion of the integrand of
Eq.~(\ref{intrepr2}) in small parameters with subsequent interchange of
summation and integration over $k$ coincides with the series which is
obtained when we {\it formally} expand the integrand of the
original integral in small parameters, using a formula analogous
to Eq.~(\ref{ss}), interchange  summation and integration, and
rewrite the integrals of the obtained series in Feynman
parametrization.
   We thus conclude that the IR regular part of the original integral can be
obtained by expanding the integrand in small parameters and interchanging
summation and integration over loop momenta.
   In practical calculations of the IR regular parts of loop integrals
it is convenient to reduce the loop integrals to integrals over
(Feynman/Schwinger) parameters, expand the integrand in Lorentz-invariant
small expansion parameters (small masses and Lorentz-invariant combinations of
external momenta and large masses), and interchange  integration
and summation.

\section{applications}
    As a check and application of our formulation
of the IR regularization we have explicitly verified for all integrals
of pion-nucleon scattering of Ref.~\cite{Becher:2001hv} (to the
order which is needed for the accuracy of calculations of that
work) that, by expanding the integrands in small parameters and
changing the order of summation and loop integration, one
reproduces the IR regular parts of these integrals.
   Although the original integrals of BChPT do not contain infrared
divergences, the IR regular parts as well as the IR singular parts separately
contain such divergences.\footnote{Note that, using dimensional regularization,
IR divergences are also parametrized as $1/(n-4)$ poles.}
   In the approach of Becher and Leutwyler these divergences  of
both parts are absorbed in counterterms. (In fact they exactly
cancel each other and hence do not give any contributions in
couterterms.)
   In our formulation the IR regularized integrals are
obtained by subtracting the IR regular parts, from which the IR
divergences are removed beforehand, from the full expressions of
the integrals.
   Clearly our expressions of the IR regularized
integrals coincide with the results of the Becher-Leutwyler
approach.

   It is straightforward to apply our formulation of IR
regularization to diagrams with multiple nucleon lines. We have
checked that our approach reproduces the results of
Ref.~\cite{Goity:2001ny} for diagrams with two nucleon
propagators.
   As an illustration let us consider the following
integral:\footnote{Our notations differ from those of
Ref.~\cite{Goity:2001ny}.}
\begin{equation}
I_{NN\pi}(P_1,-P_2,0)=i \int \frac{d^n k}{(2\pi)^n}
\frac{1}{\left[
(k+P_1)^2-m^2+i 0^+ \right] \left[ (k-P_2)^2-m^2+i 0^+\right]
\left[ k^2-M^2+i 0^+\right]}.
\label{glpsint}
\end{equation}
   Using Feynman parametrization and performing loop
momenta integration one can write $I_{NN\pi}$ as
\cite{Goity:2001ny}
\begin{equation}
I_{NN\pi}(P_1,-P_2,0)=\frac{1}{(4\pi)^{n/2}} \ \frac{\Gamma\left(
3-n/2 \right)}{2} \int_0^1 dz \ z \int_{-1}^1 dw {\left[ C(w,z)-i
0^+\right]^{(n/2)-3}},
\label{glpsintfp}
\end{equation}
where $$C(w,z)=(1-z) M^2+z m^2-z^2 (1-w^2)
\frac{(P_1+P_2)^2}{4}-\frac{z (1-z)(P_1^2+P_2^2)}{2}-\frac{w z
(1-z) (P_1^2-P_2^2)}{2}.$$
   Following Ref.~\cite{Goity:2001ny} we define the IR regular part of the
integral $I_{NN\pi}(P_1,-P_2,0)$ as
\begin{equation}
R_{NN\pi}(P_1,-P_2,0)=\frac{1}{(4\pi)^{n/2}} \ \frac{\Gamma\left(
3-n/2 \right)}{2} \int_1^\infty dz \ z \left( \int_{-\infty}^{-1}
dw +\int_{1}^\infty dw \right){\left[ C(w,z)-i
0^+\right]^{(n/2)-3}}. \label{glpsintfpR}
\end{equation}
   To calculate $R_{NN\pi}$ we expand the integrand of
Eq.~(\ref{glpsintfpR}) in powers of $M^2$, $4 m^2-(P_1+P_2)^2$,
$P_1^2-m^2$, and $P_2^2-m^2$ and interchange integration and
summation \cite{Goity:2001ny}.
   Doing so we obtain a series, the coefficients of which are proportional
to the integrals
$$I_{ij}= \int_1^\infty dz \ (z^2)^{(n/2)-3} z^{1+i} \left(
\int_{-\infty}^{-1} dw +\int_{1}^\infty dw \right) (w^2)^{(n/2)-3}
w^{j},$$
where $i$ and $j$ are integers.
   Again, the integrals $I_{ij}$ are calculated by analytical continuation
from the domain of $n$ in which they converge, leading to
\begin{equation}
I_{ij}=\frac{1+(-1)^j}{(n-4+i)(n-5+j)}.
\label{Iij}
\end{equation}
   On the other hand, in our approach we identify the IR regular part
of $I_{NN\pi}$ by expanding the integrand in Eq.~(\ref{glpsintfp})
in powers of small parameters [$M^2$, $4 m^2-(P_1+P_2)^2$,
$P_1^2-m^2$, and $P_2^2-m^2$] and interchanging summation and
integration over Feynman parameters.
   This leads to exactly the same expansion that we obtained above for
$R_{NN\pi}(P_1,-P_2,0)$, but instead of the integrals $I_{ij}$ we now have
$$J_{ij}= \int_0^1 dz (z^2)^{(n/2)-3} z^{1+i} \int_{-1}^{1} dw
(w^2)^{(n/2)-3} w^{j},$$
which we calculate by analytically continuing from the domain of $n$ in which
they converge:
\begin{equation}
J_{ij}=\frac{1+(-1)^j}{(n-4+i)(n-5+j)}. \label{Jij}
\end{equation}
   Clearly, in analogy to the one-nucleon sector our formulation of
IR regularization reproduces the results of Ref.~\cite{Goity:2001ny} for
diagrams involving two nucleon lines.

   Recently, we have shown \cite{Fuchs:2003sh} that, within
the EOMS renormalization scheme, one can set up a consistent power
counting in the effective field theory with (axial) vector mesons
included explicitly.
   Analogously we could apply the IR regularization in our formulation.
   When treating vector mesons in the antisymmetric tensor field representation
and analyzing the diagrams contributing to the electromagnetic form
factors of the nucleon up to and including ${\cal O}(q^4)$, we observe that
in Ref.~\cite{Kubis:2000zd} all relevant loop diagrams have actually been
taken into account.
   This is due to the fact that the integrals involving only vector meson
and nucleon propagators vanish in IR regularization.

    Finally, we have also applied our formulation of the IR regularization
to the integral considered in Ref.~\cite{Bernard:2003xf} in the
context of including the $\Delta$ resonance:
\begin{equation}
I_{0\pi}(-p,0)=i\int\frac{d^nk}{(2\pi)^n}\frac{1}{
\left[ (k-p)^2+i
0^+ \right]\left[ k^2-M^2+i 0^+\right]}.
\label{bhmint}
\end{equation}
   For $p^2\gg M^2$ the chiral dimension of this integral
should be $Q^{n-2}$ \cite{Bernard:2003xf}.
   An explicit calculation of the integral $I_{0\pi}$ results in
\begin{equation}
I_{0\pi}(-p,0)=
-\frac{M^{n-4}}{(4\pi)^{n/2}}\frac{\Gamma(2-n/2)\Gamma(n/2-1)}{\Gamma(n/2)}
\
F\left(1,2-n/2;n/2;\frac{p^2+i 0^+}{M^2}\right),
\label{bhmintcalc}
\end{equation}
   where $F(a,b;c;z)$ is the hypergeometric function
\cite{as}.
   For $p^2>M^2$ we rewrite Eq.~(\ref{bhmintcalc}) as
\begin{eqnarray}
I_{0\pi}(-p,0)&=&\frac{M^{n-2}}{(4\pi)^{n/2}}\frac{\Gamma[1-n/2]}{p^2}
\
F\left(1,2-n/2;n/2;\frac{M^2}{p^2+i 0^+}\right)\nonumber\\
&&-
\frac{(-p^2-i 0^+)^{(n/2)-2}}{(4\pi)^{n/2}} \ \frac{\Gamma(2-n/2)
\Gamma(n/2-1)\Gamma(n/2-1)}{\Gamma(n-2)}
\ \left(1-\frac{M^2}{p^2+i 0^+}\right)^{n-3}.\nonumber\\
\label{bhmintrewr}
\end{eqnarray}
   Analogously to Ref.~\cite{Becher:1999he} we identify the
first term in Eq.~(\ref{bhmintrewr}), which, for noninteger values of $n$,
is proportional to a noninteger power of $M$, as the IR singular
part and the second term as the IR regular part.
   The IR singular part satisfies the power counting and would generate
an infinite number of terms if the function multiplying $M^{n-2}$ were
expanded in powers of $M^2$.
   This differs from the result of Ref.~\cite{Bernard:2003xf}, where
only the first term of such an expansion was identified as the IR
singular part of $I_{0\pi}(-p,0)$.

   In analogy to the self-energy integral considered above, it is
straightforward
to check explicitly that, if one expands the integrand of Eq.~(\ref{bhmint})
in powers of $M^2$ and interchanges integration and summation, one exactly
reproduces the expansion of the second (regular) term of
Eq.~(\ref{bhmintrewr}) in powers of $M^2/p^2$.

   The integral $I_{0\pi}(-p,0)$ has an imaginary part for $p^2>M^2$
in both definitions, ours and that of Ref.~\cite{Bernard:2003xf}
[provided that in this work the same boundary condition has tacitly been
assumed as in Eq.~(\ref{bhmint})], which is included in the IR regular
part.
   This imaginary part is given by
$$
-\frac{1}{16\pi}(1-M^2/p^2)
$$
in $n=4$ dimensions and violates the power counting.
   Therefore, although the regular part is analytic in $M^2$ and consequently
its real part can be absorbed by counterterms of the Lagrangian, the
imaginary part cannot be altered.
   As a result there exists no subtraction scheme within which the
renormalized version of $I_{0\pi}(-p,0)$ would satisfy the power counting.
   However, from this observation one should {\it not} draw the conclusion
that there is no consistent power counting in a manifestly Lorentz-invariant
formulation of BChPT with spin 3/2 particles included explicitly.
    Rather, as already pointed out in Ref.\ \cite{Bernard:2003xf},
the integral $I_{0\pi}(-p,0)$ occurs when the spin 3/2 particle
propagator is decomposed using projection operators
and the apparent puzzle disappears once the results for this
decomposition are put together.

\section{\label{summary} Summary and Conclusions}

   We have reformulated the IR regularization of Becher and
Leutwyler \cite{Becher:1999he} in a form analogous to our EOMS
renormalization scheme of Ref.~\cite{Fuchs:2003qc}.
   Within this (new) formulation the subtraction terms are found by
expanding the integrands of loop integrals in powers of small
parameters (small masses and Lorentz-invariant combinations of
external momenta and large masses) and subsequently exchanging the order of
integration and summation.
   Isolating the infrared divergences from these terms and subtracting them
from the original relativistic loop integral one obtains the IR
regularized expression of the integral.
   Within our formulation of IR regularization one does not
necessarily need to use dimensional regularization.
   The renormalized results are regularization-scheme independent, provided
that the infrared divergences are properly handled, and one could thus also
use, e.g., Pauli-Villars regularization.
   One advantage of the new formulation of IR regularization is that
it can be easily applied to diagrams with an arbitrary number of
propagators with various masses (e.g., resonances) and/or diagrams
with several fermion lines as well as to multi-loop diagrams.

\acknowledgments The work of S.~Scherer  was supported by the
Deutsche Forschungsgemeinschaft (SFB 443).
   J.~Gegelia acknowledges the support of the Alexander von Humboldt
Foundation.

\end{document}